\documentclass[useAMS,usenatbib]{mn2e}
\usepackage{graphicx}
\usepackage{mathrsfs}

\title[]{Atmospheric optical-turbulence at Roque de los Muchachos Observatory: database and recalibration of the generalized-SCIDAR data}

\author[Garc\'{\i}a-Lorenzo et al.]{B. Garc\'{\i}a-Lorenzo$^{1,2}$\thanks{E-mail:
bgarcia@iac.es}, \& J.J. Fuensalida$^{1,2}$ \\
1 Instituto de Astrof\'{\i}sica de Canarias, C/Via Lactea S/N, 38305-La Laguna, Tenerife, Spain \\
2 Dept. Astrof\'{\i}sica, Universidad de La Laguna, C/ Astrof\'{\i}sico Francisco S\'anchez, E-38205 Tenerife, Spain \\}
\begin{document}

\date{Accepted ..... Received .....; in original form .....}

\pagerange{\pageref{firstpage}--\pageref{lastpage}} \pubyear{2004}

\maketitle

\label{firstpage}

\begin{abstract} 

We present the largest database so far of atmospheric optical-turbulence profiles (197035 individual C$_N^2(h)$) for an astronomical site, the Roque de los Muchachos Observatory (La Palma, Spain). This C$_N^2$(h) database was obtained through generalized-SCIDAR observations at the 1 meter Jacobus Kapteyn telescope from Febrary 2004 to August 2009, obtaining useful data for 211 nights. The overestimation of the turbulence strength induced during the generalized SCIDAR data processing has been analysed for the different observational configurations. All the individual C$_N^2$(h) have been re-calibrated to compensate the introduced errors during data treatment following \citep{avila09}. Comparing results from profiles before and after the recalibration, we analyse its impact on the calculation of relevant parameters for adaptive optics.
\end{abstract}

\begin{keywords}
Atmospheric effects ---- instrumentation: adaptive optics ---- Site Testing
\end{keywords}

\section{Introduction}

The atmospheric optical turbulence is an effect that acts on the propagation of light waves through the atmosphere. Its origin is on random variations of the refractive index associated to temperature fluctuations. Atmospheric optical turbulence drastically affects to astronomical observations, limiting the capabilities of ground-based telescopes. The refractive index structure parameter, C$_N^2$, constitutes a quantitative measure of the atmospheric optical turbulence strength \citep{tatarskii71}[e.g.] which depends on the position. C$_N^2$ as a function of the altitude is commonly referred to as optical turbulence profile, being a relevant variable in the definition of adaptive optics systems in astronomy. The SCIntillation Detection And Ranging (SCIDAR) technique is an efficient technique to measure the optical turbulence profiles in astronomical observatories. SCIDAR is based on the analysis of the intensity distribution (scintillation patterns) at the pupil plane (classical-SCIDAR) or at a virtual plane (generalized-SCIDAR) of a telescope when observing a double-star. Atmospheric turbulence profiles are derived through the inversion of the normalized autocovariance of a large number of scintillation patterns. The SCIDAR technique, in its classical or generalized versions, has been extensively explained in previous papers \citep{vernin73, rocca74, fuchs94, avila97, kluckers98, johnston02}. The classical-SCIDAR \citep{vernin73, rocca74} is not sensitive to turbulence at the observatory level while generalized-SCIDAR \citep{fuchs94, avila97, kluckers98, johnston02} is able to measure this turbulence by locating the analysis plane in a virtual position a few kilometers below the pupil plane. Generalized SCIDAR has been extensively used in astronomical sites to study the atmospheric optical turbulence for applications in the development of adaptive optics systems and site characterization \citep{avila97, fuchs98, kluckers98, avila98, kern00, prieur01, avila01, wilson03,avila04,tokovinin05,garcia06,egner07,vernin07, salida08, garcia09, chun09, masciadri10, dali10, mohr10, garcia11}. From the early experimental implementations of the technique, it was noted that generalized-SCIDAR data processing leads to an overestimation of the optical turbulence strength \citep{johnston02} that was assumed as negligible at any altitude. \cite{avila09} demonstrated that this overestimation might be negligible or relevant strongly depending on the selected observational parameters -- namely telescope diameter, double-star angular separation and analysis plane conjugation altitude. An analytical expression to calculate the actual errors induced during generalized-SCIDAR data processing as well as a procedure for the correct recalibration of C$_N^2$ profiles derived from generalized-SCIDAR observations were also provided in \cite{avila09}. This procedure has been already applied to re-calibrate atmospheric optical turbulence profiles retrieved in Mt Graham \citep{masciadri10}, El Teide \citep{garcia11}, and San Pedro M\'artir \citep{avila11} observatories.

The atmospheric optical turbulence monitoring programme at the Roque de los Muchachos Observatory (ORM hereafter) started in 2004. The classical data processing \citep{kluckers98}[see e.g.] was performed to retrieved the C$_N^2$ profiles from the generalized-SCIDAR observations assuming negligible errors induced in the data treatment. Results derived from the C$_N^2$ profiles in the ORM database were already published \citep{castro09, garcia09b, garcia07, salida07, salida04a, salida04b} before the \cite{avila09} work. In this paper, we present the database of atmospheric optical turbulence profiles recorded at the ORM through generalized-SCIDAR observations, the largest database of C$_N^2$(h) for an astronomical site that has been published so far. We also perform the recalibration of the C$_N^2$ profiles in the ORM database to compensate for the errors introduced during data processing. We analyze the implications in the statistical results derived from this database before and after the recalibration. Section \S2 presents the database of C$_N^2$ profiles derived from generalized-SCIDAR measurements at the ORM. We also calculate the impact of the data processing error on retrieved C$_N^2$ profiles and re-calibrate the full database following the proposed procedure in \cite{avila09}. Section \S3 analyzes the implications of the C$_N^2$ database recalibration on results derived from profiles. Conclusions are summarized in section \S4.

\section{THE DATA}

The ORM is located at an altitude of $\sim2396$ meters above sea level ({\it asl} hereafter), at latitude $28^0 46'$ N and longitude $17^0 53'$ W on the island of La Palma (Canary Islands, Spain). This astronomical site was one of the final candidates to locate the European Extremely Large Telescope (42m-EELT). A monitoring program of the atmospheric turbulence structure at the ORM began in 2004 using the generalized-SCIDAR technique. The 1-m Jacobus Kapteyn Telescope was used in combination with the Cute-SCIDAR instrument \citep{ho04, salida04c} developed at the Instituto de Astrof\'{\i}sica de Canarias (Tenerife, Canary Islands, Spain). Each detector pixel of Cute-SCIDAR instrument covers a square 1.935 cm in size on the 1-m Jacobus Kapteyn Telescope pupil. The generalized-SCIDAR data were processed using the traditional procedure \citep{kluckers98}[see e.g.] of deriving the normalized autocovariance from a series of scintillation patterns (1000 images in ORM case). The autocovariance peaks allow the determination of C$_N^2$(h) using a numerical inversion.
The C$_N^2$(h) systematic campaigns at ORM were carried out from February 2004 to October 2006 and from January 2008 to August 2009 with a frequency of about 4-6 nights per month. Useful generalized-SCIDAR observations were obtained in 211 nights during these campains and 197035 individual C$_N^2$ profiles constitute the database of turbulence profiles at ORM (see table \ref{tab1}), the largest C$_N^2$ database published until now. The dome and mirror turbulence contribution was removed from all the profiles using the procedure in \cite{salida08}.
\begin{table*}
\caption{Distribution of nights in the periods from February 2004 to October 2006 and from January 2008 to May 2009 in which useful generalized-SCIDAR measurements were obtained at the Roque de los Muchachos Observatory. The number of individual turbulence profiles recorded each night is also included.}
\begin{tabular}{l|r|rrrrrrrrrrrr||lr}
           & {\bf Year} &            &            &            &            &            &            &            &            &            &           &            &           &  \vline & {\bf Total}  \\ 
{\bf Month}      &      & {\bf Jan}  & {\bf Feb}  & {\bf Mar}  & {\bf Apr}  & {\bf May}  & {\bf Jun}  & {\bf Jul}  & {\bf Aug}  & {\bf Sep}  & {\bf Oct} & {\bf Nov}  & {\bf Dec} &  \vline &            \\  \hline  
{\bf Profiles}   & 2004 & ---        & 2323        & 5942        & 9586       & 5953       & 5702       & 7530       & 8186       & 8867       & 1260     &      3847  & 3191      &  \vline &  62387      \\
{\bf Nights  }   &      & ---        & 3           &  6          & 8          & 6          & 5          &  7         & 7          & 9          & 2        &        4   &  4        &  \vline &     61      \\ \hline
{\bf Profiles}   & 2005 &  10347     & 380         &  1029       & 4450       & 4039       & 5116       & 4313       & 1337       & 1723       & 2715     &     1035   &  2123     &  \vline &  38607     \\
{\bf Nights  }   &      &  7         & 1           &  3          & 4          &  4         &  7         &  4         &   3        &  3         & 3        &        2   &     2     &  \vline &     43      \\ \hline
{\bf Profiles}   & 2006 &  2460      & 2478        &  1640       & 2710       & 4363       &  0         &  3203      & 1933       &  1329      & 275      &        0   &     0     &  \vline &  20391     \\
{\bf Nights  }   &      &  2         & 2           &  2          & 3          &  4         &  0         &  3         &   2        &  1         & 1        &        0   &     0     &  \vline &     20    \\ \hline
{\bf Profiles}   & 2008 &  2860      & 961         &  5462       & 671        & 3200       &  7150      &  4818      & 5818       &  3257      & 2088     &     3236   &  3361     &  \vline &  42882    \\
{\bf Nights  }   &      &  4         & 1           &  6          & 2          &  3         &  7         &  8         &   8        &  4         & 2        &        5   &     3     &  \vline &     53     \\ \hline
{\bf Profiles}   & 2009 &  8287      & 4048        &  4805       & 3316       &  3365      &  3162      & 2170       & 3615       & ---        & ---      &  ---       &  ---      &  \vline &  32768  \\
{\bf Nights  }   &      &  8         & 4           &  4          & 4          &  4         &  3         &  3         &   4        & ---        & ---      &     ---    &  ---      &  \vline &     34   \\ \hline \hline
{\bf Profiles}   & {\bf Total} & 23954  & 10190     & 18878       &  20733     &  20920     & 21130      &  22034     &  20889     & 15176      &  6338    &     8118   &   8675    &  \vline & 197035  \\
{\bf Nights  }   &      &  21        & 11           &  21         & 21         & 21         & 22         &  25        & 24         & 17         & 8        &       11   &      9    &  \vline &    211  \\ \hline
\end{tabular}
\label{tab1}
\end{table*}

  A set of 19 double-stars (see table \ref{stars}) were selected to carry out the generalized-SCIDAR observations at the ORM. The double-stars selection was based on :  (1) apparent magnitude of the primary star brighter than 6.5 and double-star magnitude difference smaller than 2.5, in order to garantee an appropiate signal-to-noise in the autocovariance peaks; (2) double-star angular separations in the range from 4.2 to 10 arcsec, to ensure that the SCIDAR maximum altitude is high enough to detect all turbulent layers. After a few months of operations showing that most of the turbulence was concentrated in low-altitude layers, we increase the range of double-star angular separation to 16.5 arcsec to better sample low-altitude turbulence structure; (3) double-star declination in the range between 2 and 56 degrees, allowing generalized SCIDAR measurements at zenith angles shorter than 30$^0$; and (4) a variety of double-star right ascension, to allow a full monitoring of the atmospheric turbulence structure along seasons. The selected double-star systems allow to retrieved C$_N^2$(h) with vertical resolutions  \citep{avila98}[see e.g.] at ground ($\Delta H$(0)) that ranges from $\sim$195 to $\sim$1314 meters. The typical $\Delta H(0)$ for Generalized SCIDAR observations is about 1000 meters. The observations carried out at ORM are split in two groups according to $\Delta H(0)$. Hereafter Generalized SCIDAR observations obtained with a $\Delta H(0)$ lower and higher than 500 meters will be referred to as high- and low-resolution modes, respectively. Generalized SCIDAR data recorded using high vertical resolution are limited to turbulence structures up to $\sim17$ km (only $\sim$17\% of the profiles were obtained in high-vertical resolution mode), while data obtain in low vertical resolution ($\Delta H(0)>$ 500 meters) reaches well above 20 km at the 1-m Jacobus Kapteyn telescope (constituting $\sim$83\% of the profiles in the ORM database).

\begin{table*}
\caption{Main parameters of the double-stars used in the generalized-SCIDAR observations recorded at the Roque de los Muchachos Observatory. m$_1$ designates the brightest star magnitude in the V band and $\Delta$m is the difference in magnitude between both stars. $\theta$ corresponds to the angular separation between both stars. The different positions of the analysis plane ( {\it d}) respect to the pupil plane used with each double-stars system is also summarized, including the percentage of turbulence profiles (C$_N^2$(h)) recorded in each configuration. The average percentage error induced during the generalized-SCIDAR data processing for each configuration is also included. Uncertainties
 indicate only the standard deviation of the averaged errors in altitude.}
\begin{tabular}{l|cccccccc}\hline
{\bf Double-star} & $\alpha_a$             & $\delta_a$             & m$_1$ & $\Delta$m & $\theta$            & d     & Percentage & Average   \\  
                  &   (J2000)              &    (J2000)             &        &     & (arcsec)  &  (km)                 & of profiles & error (\%)  \\ \hline
{\bf BS~282  }    &       01$^h$ 00$^m$ 03$^s$   & +44$^o$ 42$'$ 40.1$''$   &   5.70        &   0.34       &  7.8        &    4   &  0.25  & 41.47$\pm$20.03 \\

{\bf BS~546  }    &       01$^h$ 53$^m$ 31$^s$  & +19$^o$ 17$'$ 38.6$''$    &   4.64       &    0.08      &   7.5        &   3    &  8.60  & 31.29$\pm$18.66  \\
                  &                       &                          &               &              &                    &  4    &  4.42  & 38.47$\pm$19.13   \\ 

{\bf BS~628  }          & 02$^h$ 10$^m$ 53$^s$  & +39$^o$ 02$'$ 22.0$''$    &   5.63       &    0.41      &   16.7        &   2   &  0.07  & 42.70$\pm$20.18  \\
 
{\bf BS~1821 }          & 05$^h$ 29$^m$ 16$^s$  & +25$^o$ 09$'$  00.7$''$   &  5.47       &     0.32       &  4.8        &  4     &   3.70 & 15.53$\pm$4.13  \\ 
                  &                        &                         &               &              &                    &  5    &   1.61 & 20.29$\pm$4.76   \\ 

{\bf BS~1847 }          & 05$^h$ 32$^m$ 14$^s$  & +17$^o$ 03$'$  29.2$''$   &  5.46       &     0.09       &  9.6         & 4     &  0.42 & 45.53$\pm$20.36  \\

{\bf BS~1879 }          & 05$^h$ 35$^m$ 08$^s$   & +09$^o$ 56$'$ 02.9$''$  &   3.30        &  2.07         & 4.4             &  4     &   0.75   & 15.55$\pm$4.53 \\
                  &                       &                          &               &              &                       &  5     &    1.13  & 18.57$\pm$5.04  \\ 

{\bf BS~2784 }          & 07$^h$ 22$^m$ 52$^s$   & +55$^o$ 16$'$ 53.0$''$  &   5.78        &  1.08         & 15.0             &    2  &   0.26   & 41.65$\pm$21.95    \\ 

{\bf BS~2890 }          & 07$^h$ 34$^m$ 35$^s$   & +31$^o$ 53$'$ 18.5$''$  &   1.93        &  1.04         & 4.4             &    5   &   5.33  & 18.34$\pm$ 4.59      \\ 

{\bf BS~3617 }          & 09$^h$ 07$^m$ 27$^s$   & +22$^o$ 58$'$ 52.0$''$  &   6.40        &  0.62         & 7.6             &    3   &   0.12  & 32.26$\pm$19.78      \\ 

{\bf BS~4057 }          &  10$^h$ 19$^m$ 58$^s$  & +19$^o$ 50$'$ 28.5$''$   &   2.61        &   0.86       &  4.6         &  4        &   4.22   & 15.33$\pm$4.13     \\ 
                 &                       &                          &               &              &                    &  5         &   12.59  & 18.46$\pm$4.51      \\ 

{\bf BS~4259}           & 10$^h$ 55$^m$ 36$^s$  &  +24$^o$ 44$'$ 59.2$''$   &   4.50       &   1.95       &  6.5          &  4        &   1.85   & 29.87$\pm$14.36       \\ 

{\bf BS~5054 }          & 13$^h$ 23$^m$ 55$^s$  &  +54$^o$ 55$'$ 31.3$''$   &   2.27       &   1.68       &  14.5         &  2        &   2.42   & 41.77$\pm$23.78          \\ 

{\bf BS~5329 }          & 14$^h$ 13$^m$ 29$^s$  &  +51$^o$ 47$'$ 25.0$''$   &   4.50       &   2.19       &  13.4         &  2        &   0.12   & 41.04$\pm$25.09      \\ 

{\bf BS~5475  }         & 14$^h$ 40$^m$ 43$^s$  & +16$^o$ 25$'$ 05.9$''$   &    4.91       &    0.94      &   5.6         &  4      &   7.15    & 21.33$\pm$7.66  \\

{\bf BS~5789 }          & 15$^h$ 34$^m$ 48$^s$  &  +10$^o$ 32$'$ 20.7$''$   &   3.80       &    0.00       &  4.3        &  5         &  6.28   & 18.18$\pm$4.42     \\
                  &                       &                         &               &              &                    &  6         &   4.34  & 21.05$\pm$4.47    \\ 

{\bf BS~5834 }          & 15$^h$ 39$^m$ 22$^s$  &  +36$^o$ 38$'$ 08.9$''$   &   5.00       &   0.96       &  6.3          &  3        &  0.89   & 20.38$\pm$9.70        \\

{\bf BS~6730 }          &  18$^h$ 01$^m$ 30$^s$  & +21$^o$ 35$'$ 44.8$''$   &   4.96        &   0.22       &  6.2         &  3       &  7.46   &  19.82$\pm$9.03      \\
                  &                        &                         &               &              &                    &  4        &  11.95  & 25.86$\pm$11.24           \\ 

{\bf BS~6781 }          &  18$^h$ 07$^m$ 49$^s$  & +26$^o$ 06$'$ 05.0$''$   &   5.86        &   0.04       &  14.3        &  2        &  0.52   & 38.78$\pm$20.37        \\

{\bf BS~7948  }         &  20$^h$ 46$^m$ 39$^s$   & +16$^o$ 07$'$ 27.4$''$  &   4.27        &   0.87      &   9.6        &  2        &   2.95  &  31.78$\pm$22.62       \\ 
                  &                        &                         &               &              &                    &  3        &  10.60   & 40.26$\pm$22.85 \\ \hline

\end{tabular}
\label{stars}
\end{table*}

\subsection{Recalibration to compensate the error induced in generalized-SCIDAR data processing}

\begin{figure*}
\centering
\includegraphics[scale=0.20]{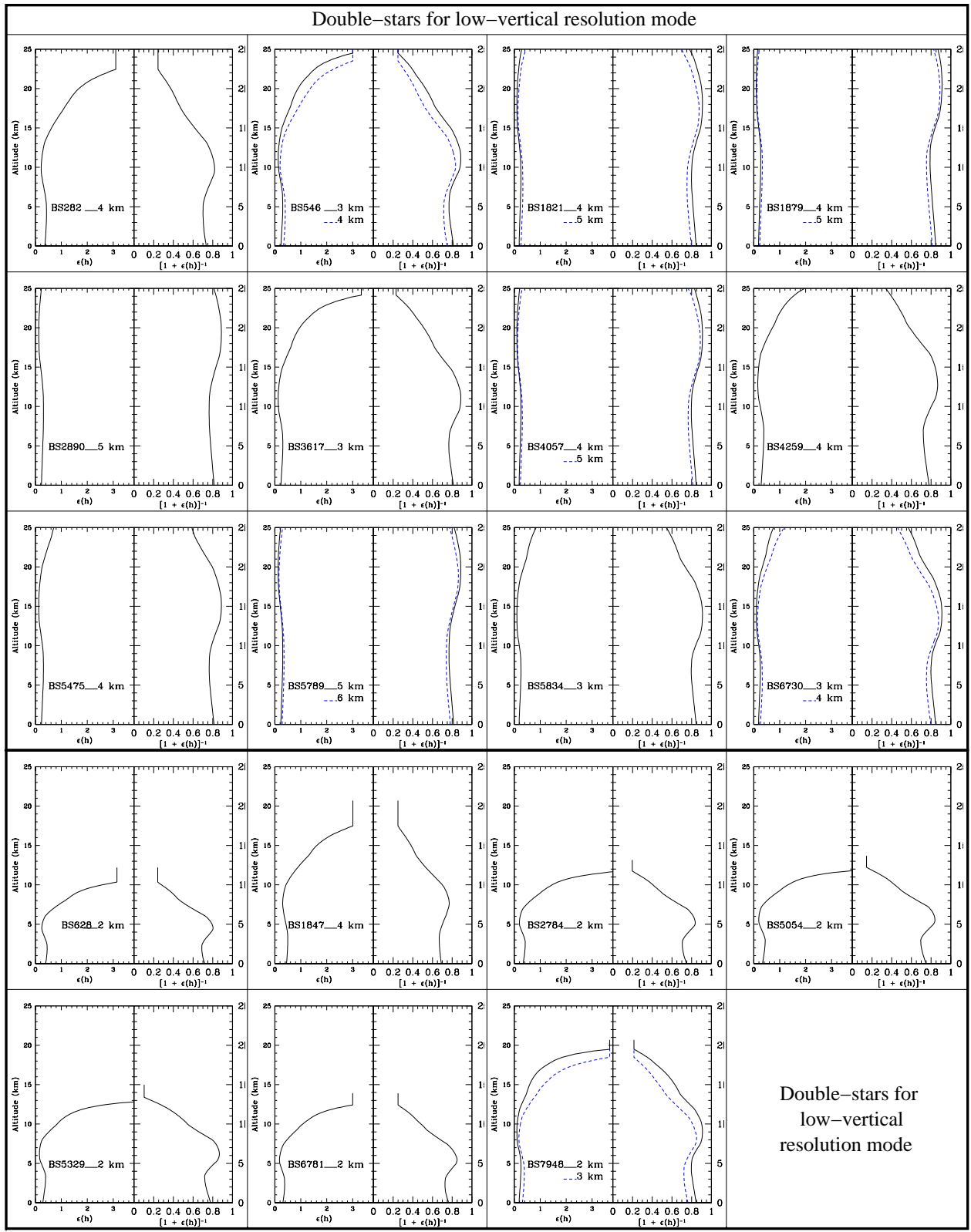}
 \caption{Vertical profiles of the relative error induced and correction factor to be applied to the measured C$_N^2$ values derived
 from generalized-SCIDAR at the 1-m Jacobus Kapteyn Telescope for all the double-stars and analysis plane combinations used to recorded 
the data at the Roque de los Muchachos Observatory. In each plot, left-pannel corresponds to the vertical distribution of the relative
error induced during the generalized-SCIDAR data processing and right-pannel is the vertical distribution of the multiplying factor.
 Different colors styles in each plot indicate different positions of the analysis plane used with the same double-star as indicated in each plot. The upper twelve plots correspond to the double-stars providing vertical resolutions at ground level larger than 500 meters (low-resolution mode). The other seven double-stars give $\Delta H(0) <$ 500 meters, that labelled high-resolution mode.  }
\label{error_juntas}
\end{figure*}

Due to the shift between the pupil footprints of the two stars on the detector (see Fig. 1 in \cite{garcia11}) during generalized-SCIDAR observations, the derived C$_N^2$ intensities are indeed an overestimation of the actual turbulence strength \citep{avila09}. This overstimation is induced during data processing and strongly depends on the selected double-stars and the analysis planes combinations as well as on the telescope diameter. Table \ref{stars} shows the different observational configurations of the generalized-SCIDAR observations carried out at the 1-m Jacobus Kapteyn telescope and, following equations in \cite{avila09}, we have calculated the actual error $\epsilon(h)$ induced in the retrieved C$_N^2$ profiles for each of these 26 different configuration (see Fig \ref{error_juntas}). The turbulence strength at any altitude should be multiplied by a factor 1/(1+$\epsilon(h)$) to compensate the error induced during the generalized-SCIDAR data processing. Table \ref{stars} includes the average 1/(1+$\epsilon$) factor and its standard deviation, as reference of an average error induced by each configuration. Figure \ref{error_juntas} clearly shows that these errors are not negligible for most of the used observational configurations at the ORM.

The derived turbulence strength at the observatory level ($\sim 2400$ m) after the generalized-SCIDAR data proccessing is overestimated in a percentage factor that ranges from $\sim11$\% to $\sim34$\%. The C$_N^2$ profiles recorded in high-vertical resolution mode are the most affected (see table \ref{stars} and Fig. \ref{error_juntas}) by the error introduced during generalized-SCIDAR data processing. A 45\% overstimation between the actual and the retrieved turbulence strength at 10 km {\it asl} is found in average for high-resolution profiles, reaching a maximum error of 72\%. Only 17\% of the profiles in the ORM were obtained in high-vertical resolution mode. For low-vertical resolution configurations, errors in C$_N^2$ profiles up to 10 km do not exceed 28\% in any case, with a median overstimation of 18.5 \%. At 18 km {\it asl}, turbulence strength in profiles retrieved using low-resolution mode presents a $\sim20$\% overestimation in average of the turbulence strength.

Following \cite{avila09}, we have re-calibrated the full database of C$_N^2$ profiles obtained at ORM multiplying the retrieved C$_N^2$ strengths at any altitude by the calculated factor 1/(1+$\epsilon$(h)) for each particular double-star and analysis plane combination.

\section{Implications on statistical profiles and parameters derived}

\begin{figure*}
\centering
\includegraphics[scale=0.40]{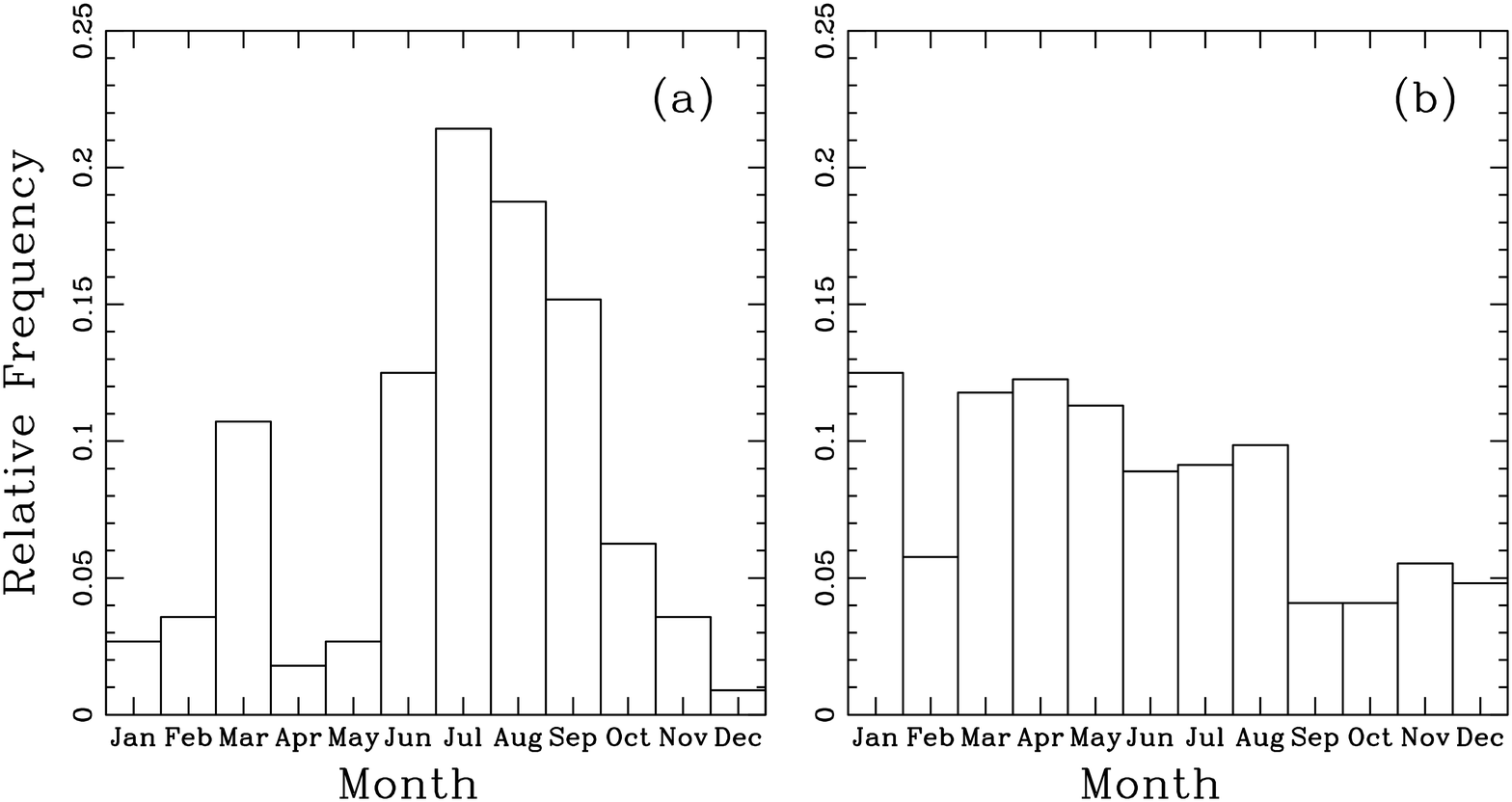}
 \caption{Relative frequency along the months of the useful nights in which generalized SCIDAR observations were obtained using: (a) high-vertical resolution mode ( $\Delta H< 500$ meters), and (b) low-vertical resolution mode ( $\Delta H> 500$ meters). }
\label{relfre}
\end{figure*}

Generalized SCIDAR observations at ORM have been taken using high-, and low-resolution modes. From Fig. \ref{error_juntas} is clear that errors are more important in profiles obtained in high-resolution mode than those derived from low-resolution observations. In order to have a view of the implications of these errors, we have derived from these profiles some statistical parameters relevant for adaptive optics purposes, namely total seeing ($\epsilon$), boundary layer (B-L) contribution to the seeing ($\epsilon_{B-L}$), free atmosphere (F-A) contribution to the seeing ($\epsilon_{F-A}$), and isoplanatic angle ($\theta_0$). The derived values for these parameters before and after the recalibration of the turbulence profiles are presented in table \ref{stat_high_low}. It is important to note that the boundary layer is not clearly defined in low-vertical resolution profiles because we are combining several profiles with $\Delta H(0)> 1000$ meters. Moreover, the reader should take into account at this point that parameters derived combining profiles obtained in high-resolution mode are affected by the limitation in altitude of these profiles. In addition, there is a seasonal bias in profiles obtained using high-resolution mode (see Fig. \ref{relfre}). For these reasons, statistical values for the different parameters derived from profiles obtained in high- and low-resolution mode are not directly comparable. As we already noted, the error induced during Generalized SCIDAR data processing leads an overestimation of the turbulence intensity \citep{avila09}. For this reason, the seeing values ($\epsilon$, $\epsilon_{B-L}$, $\epsilon_{F-A}$) derived before the recalibration of the profiles are larger than the corresponding values obtained from re-calibrated C$_N^2$. In the case of $\theta_0$, the error introduced by the generalized SCIDAR data processing results in an underestimation of this parameter.

\begin{table*}
\caption{Statistical values of the atmospheric parameters derived from the generalized-SCIDAR observations obtained using high-resolution ($\Delta H< 500$ meters) and low-resolution ( $\Delta H> 500$ meters) modes  .
Statistical values derived from data ignoring the errors introduced by the normalization of the
 scintillation autocovariance in the calculation of the C$_N^2$ (before) are compared to those obtained from profiles after
multiplying by the appropiate factor to compensate such errors (after).}
\begin{tabular}{ll|cc|cc|cc} \hline
                      &              &  \multicolumn{2}{|c|}{{\bf High-resolution C$_N^2$}}    & \multicolumn{2}{|c|}{{\bf Low-resolution C$_N^2$}} \\ 
                      & recalibration & {\bf before}        & {\bf after}    & {\bf before} & {\bf after} \\ \hline
{\bf Seeing ('')}     & Average      &  0.87 & 0.71   & 1.07  & 0.93     \\
                      & Median       &  0.78 & 0.63   & 0.96  & 0.84    \\
                      & $\sigma$     &  0.41 & 0.34   & 0.43  & 0.38    \\
{\bf BL ('')}         & Average      &  0.57 & 0.48   & 0.85  & 0.75     \\
                      & Median       &  0.49 & 0.41   & 0.76  & 0.67    \\
                      & $\sigma$     &  0.42 & 0.35   & 0.43  & 0.38    \\
{\bf FA('')}          & Average      &  0.52 & 0.41   & 0.49  & 0.42     \\
                      & Median       &  0.45 & 0.35   & 0.43  & 0.37    \\
                      & $\sigma$     &  0.25 & 0.22   & 0.25  & 0.21    \\
{\bf $\theta_0$('')}  & Average      &  1.97 & 2.96   & 2.02  & 2.39     \\
                      & Median       &  1.82 & 2.74   & 1.88  & 2.22    \\
                      & $\sigma$     &  0.81 & 1.19   & 2.35  & 2.26    \\ \hline
\end{tabular}
\label{stat_high_low}
\end{table*}

\begin{figure*}
\centering
\includegraphics[scale=0.4]{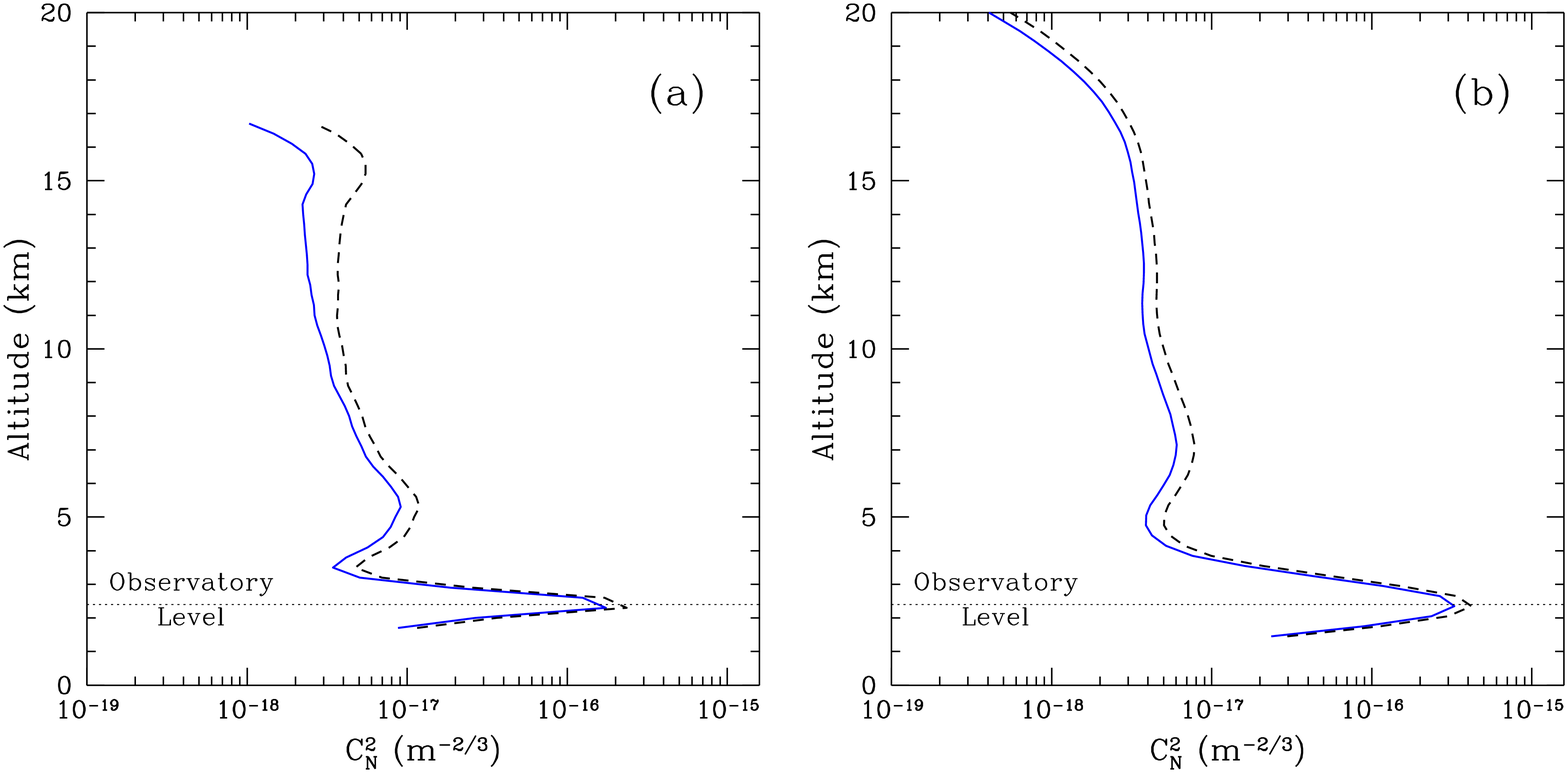}
 \caption{(a) Median C$_N^2$(h) profile obtained from the 33854 individual uncalibrated profiles (dashed-line) derived from the generalized SCIDAR observations using high-resolution modes. The solid-line corresponds to the median profile obtained through the 33854 individual profiles re-calibrated to compensate for the error induced by the normalization during data processing. (b) The same as (a) but for the 163181 individual turbulence profiles obtained in low-resolution generalized SCIDAR observations. The dome turbulence contribution was removed from individual profiles following the procedure in \citep{salida08} before combining the profiles. The horizontal dotted line indicates the observatory level ($\sim$2400 m). The horizontal axis is the turbulence strength on a logarithmic scale. The vertical axis is the altitude above sea level.  }
\label{median_profiles}
\end{figure*}

Figure \ref{median_profiles} shows the average turbulence profile derived from the combination of C$_N^2$(h) in high- (Fig. \ref{median_profiles}a), and low-resolution (Fig. \ref{median_profiles}b) modes before (dashed-line) and after (solid-line) the recalibration of the profiles. As it has been already shown in previous recalibrations \citep{avila11, garcia11}, the effect of the induced error during data processing consists of an overestimation of the turbulence strength, but following the actual turbulence structure. 

Generalized SCIDAR observations in high-resolution mode were obtained when the normalized autocovariance showed no evidence of high-altitude turbulence layers, but stratified turbulence at low level (the selection of this mode was subject to observer criteria and experience). Moreover, the use of high-resolution mode presents a seasonal bias according to Fig. \ref{relfre}. For these reasons, the median high-resolution turbulence profile (Fig. \ref{median_profiles}a) derived for ORM could not be representative of a statistical turbulence profile at this site. Fig. \ref{median_profiles}a shows the median recalibrated profile where $\sim$67\% of the detected turbulence is concentrated at the boundary layer (B-L), being 96.6\% of this turbulence located in the first 500 meters above the observatory level. A clear turbulence layer is resolved at about 5.2 km {\it asl}, that constitute about 12\% of the turbulence in the median high-resolution C$_N^2$ profile (integrating C$_N^2$(h) from 3.4 to 7 km). The turbulence upwards 7 km {\it asl} represents only 11\% of the total turbulence measured in the derived high-resolution C$_N^2$ profile. The median low-resolution C$_N^2$(h) derived for ORM (Fig. \ref{median_profiles}b) presents a smoother structure in altitude compared with the high-resolution profile. 76\% of the turbulence is concentrated in the B-L, while turbulence above 5 km represents only a $\sim$11 \% of the total turbulence. A turbulence feature appears at $\sim$7.2 km {\it asl} with a strength at the peak of about 6$\times$10$^{-18}$ m$^{-2/3}$. Any other turbulence layer is not clear in this median low-resolution turbulence profile, being the background turbulence at any altitude above 10 km always bellow 3.5$\times$10$^{-18}$ m$^{-2/3}$. There is a 2 km difference in altitude for a mid-altitude (3 km $<$ H $<$ 10 km) turbulence layer in the median high- and low-vertical resolution profiles. This difference is well-explained taking into account the seasonal variation of the turbulence structure already reported for the Canary Islands observatories \citep{garcia11,  garcia09b, garcia07, salida07}. Most of the high-vertical resolution profiles (53 \%) were recorded between June and August (see Fig. \ref{relfre}a): turbulence structure in these months is characterized by a relatively strong turbulence layer at $\sim$5-6 km {\it asl} that is stable from year to year \citep{garcia09b}. This turbulence layer evolves in altitude and strength along the year  \citep{garcia11,  garcia09b, garcia07, salida07}. The relative frequency of turbulence profiles recorded in low-vertical resolution mode ( Fig. \ref{relfre}b) has not a clear peak at any month, better smoothing the seasonal evolution of the turbulence structure above ORM.

\section{Conclusions}

The optical atmospheric turbulence structure has been monitoring since 2004 at the Roque de los Muchachos Observatory (La Palma, Canary Islands, Spain). Useful generalized SCIDAR measurements were obtained during 211 nights. The total number of individual C$_N^2$(h) profiles recorded at this site is 197035. The error induced during generalized SCIDAR data processing has been calculated, being more significant when using high-vertical resolution mode ( $\Delta H(0)<$ 500 meters). Following the procedure proposed by \cite{avila09}, we have re-calibrated the full database of turbulence profiles recorded at ORM, showing the effects of theses errors in the calculation of statistical atmospheric parameters relevant for adaptive optics. Combining the corrected turbulence profiles, we have obtained the statistical high- ( $\Delta H(0)<$ 500 meters) and low-vertical ( $\Delta H(0)>$ 500 meters) resolution turbulence profiles to have a view of the turbulence structure at ORM. The main conclusions that we have derived from this work can be summarized as follows.

\begin{itemize}
\item The generalized SCIDAR data processing leads to an overstimation of the optical turbulence strength that it is not negligible for most of the observational configurations used at Roque de los Muchachos Observatory.

\item  The error introduced during the processing of the generalized SCIDAR data can drastically affect the statistical values derived for atmospheric parameters relevant for adaptive optics (namely total seeing, boundary-layer, free-atmosphere contributions and isoplanatic angle), being as large as 50 per cent for high-altitude layers in some generalized SCIDAR configurations used at ORM.
 
\item Both the high- and low-vertical resolution profiles obtained for ORM show that most of the optical turbulence is concentrated in the first 5 km. The most intense turbulence layer is at the observatory level. A lower strength turbulence layer is detected in mid-altitude levels (4 $<$ H $<$ 8 km).
\end{itemize}

The C$_N^2$(h) set recorded at the Roque de los Muchachos Observatory constitutes the largest database of optical atmospheric turbulence profiles so far. 

\section*{Acknowledgments}

This paper is based on observations obtained at the Jacobus Kapteyn Telescope operated by the Isaac Newton Group at the Observatorio de Roque de los Muchachos of the Instituto de Astrof\'{\i}sica de Canarias.  The authors thank all the staff at the observatory for their kind support. Thanks are also due to all the observers that have recorded generalized SCIDAR data at this site (J. Castro-Almaz\'an, S. Chueca, J.M. Delgado, E. Sanroma, C. Hoegemann, M.A.C. Rodríguez-Hernández, and H. V\'azquez-Rami\'o). We also thank A. Eff-Darwich for help and useful discussions. We are grateful to the referee, Remy Avila, whose comments helped to improve this paper. This work was partially funded by the Instituto de Astrof\'{\i}sica de Canarias and by the Spanish Ministerio de Educaci\'on y Ciencia (AYA2006-13682 and AYA2009-12903). B. Garc\'{\i}a-Lorenzo thanks the support from the Ram\'on y Cajal program by the Spanish Ministerio de Ciencia e innovaci\'on.

\end{document}